# Applying CMM Towards an m-Learning Context

Preliminary Work

Muasaad Alrasheedi
Department of Electrical and Computer Engineering
University of Western Ontario
London, ON, Canada
malrash@uwo.ca

Luiz Fernando Capretz
Department of Electrical and Computer Engineering
University of Western Ontario
London, ON, Canada
lcapretz@uwo.ca

*Abstract*—In the era of m-Learning, it is found that educational institutions have onus of incorporating the latest technological innovations that can be accepted and understood widely. While investigating the important theme of fast-paced development of emerging technologies in mobile communications, it is important to recognize the extent influence of these innovations using which society can communicate, learn, access information, and, additionally, interact. In addition, the usage of mobile technology in higher education needs not only the pervasive nature of the technology but also its disruptive nature that offers several challenges while incorporation in the area of teaching and learning.

Therefore, recently, higher education institutions are looking at various ways of implementing m-Learning strategies, in order to offer solutions, which, in turn, can standardize the process of education and, additionally, replace those traditional didactic courses, focusing on m-Learning's endless benefits. Some of the benefits are: the process of learning itself could be self-paced, whereas information could be easier accessed, adding to independent, discovery-oriented learning that becomes more engaging. Applying CMM successfully to design effective incorporation strategies of m-Learning, this research targets formulation of such a maturity model by which the process of m-Learning can be more effective and efficient.

## I. INTRODUCTION

When one can recognize the highest techniques of mobiles learning that they could, in turn, evolve and assist in the experience of learning. This is true in both stand-alone and blended contexts, particularly, in those fields of m-Learning that have, in turn, emerged as such a new and valuable learning paradigm. This, indeed, become the critical focus of the research and development activities of Kukulska-Hulme et al. [1] and Engel et al. [2]. Still, there is the necessity to mention that during previous years one can easily notice many experimental researches studies that have been successfully conducted across many involved sectors simply, in order to investigate the influence of mobile technologies on the process of teaching and learning [3][4]. To make the matter clear, it is easy to find one of the most consistent results of these available studies that corresponds to the existence of many challenges and which, indeed, could impact on m-Learning adoption.

Therefore, a few higher education institutions have consequently, implemented such well-financed as well as highly visible initiatives of m-Learning that can be seen operating within practice and policy [5]. From this point, it is critical to refer that Wishart & Green [6] indicate that the most prominent challenge in the m-Learning process is the rare evaluation of the full scale of the technology of mobile technologies, namely in the higher education. One can add that a significant challenge, which is still facing most institutions of higher education, is this identification of the ideal priorities of strategy and operation for making it easy to invest in the industry of m-Learning, especially to cope with the quickly changing environment of technology [3]. This can refer to the fact that, while the results of educational process for the correspondence students are maximized at the same time the costs of the institutional processes are reduced.

In this paper, we will address such a serious gap simply by developing efficient assessment mechanisms which could be employed in order to evaluate the involved institution, especially in the terms of m-Learning adoption and implementations. There are vital question that need to be answered within this research, such as: How can we apply the technique of CMM in the process of m-Learning simply to provide better assessment and evaluation for the education institutions?

Focusing on the fact that the educational requirements of the students are highly subordinate to the requirements of technical teaching, it appears to be hard for the involved educators to facilitate the process of learning and for the teachers to acquire such hands-on experiences [2]. Generally, E-learning and m-Learning can offer valuable models of learning, through which teachers in the many involved universities can have indeed ongoing access to the available informational resources and employ these modules with care.

m-Learning is, consequently, a subtype of e-learning that can employ personal digital assistants in order to bring such information to the learners, with or without the process of accessing Internet. These complicated but comprehensive technologies can, in fact, target either or both learning enhancement and knowledge delivery, simply to build skills and knowledge [7].

This can be particularly seen as promising however, when such technological advancements can utilize both contents and methods of multi-media instructions. Therefore, it is important to notice that m-Learning can allow the involved students to access the required information according to their corresponded schedules. Moreover, it can provide additional chances to revise the needed materials [8].

Recently, most education experts have begun to assess such a utilization of the new attitudes of educational technologies. However, other studies have shown that the process of learning based techniques, such as E-learning, m-Learning, and other advanced methods, could be equal to the traditional lecture format [7]. Even so, the framework used to assess the any software application platform, including m-Learning, should not confine itself to technical & design aspects but must also include metrics determining ease of adoption and usage for users [9].

In other words, the apparent lack of such a valuable technique is the absence of such an overall framework, by which we can guide the adoption of m-Learning and, consequently, improve the process involved simply to ensure certain sustainability in enhancing the outcomes of students learning. Considering such urgent importance of m-Learning, the matter requires figuring out the best available designed software, in order to check the value of adoption. Among many available techniques, one can consider Capability Maturity Model (CMM) as an advanced software engineering discipline [10].

This paper is structured as follows: Section II provides the objectives of the research. Section III introduces the research goal. Section IV provides a literature review. Related work is presented in Section V. Section VI provides an analysis and discussion on proposed model. Finally, the conclusion is stated in section VII.

## II. RESEARCH OBJECTIVE

This paper articulates an in-depth argument regarding the validity of applying CMM to the m-Learning domain. Consequently, two possible applications of the usage of CMM will be clarified: a clear guidance that can enhance m-Learning adoption and such guidance to the corresponded adoption, along with integrating the process of E-Learning. Possible merits as well as the pitfalls of employing such an adapted version of CMM for the process of m-Learning, will be discussed at the end. Therefore, this paper attempts to answer the main question: can we creatively apply CMM in the process of m-Learning?

## III. RESEARCH GOALS

The basic goal of such valuable research is figuring out maturity model for m-Learning anywhere in the paper.

Moreover, this research aims to illustrate such basic domains and stages in the maturity model of m-Learning, in order to assist the process of assessing m-Learning's performance.

## IV. LITERATURE REVIEW

### A. Evolution of m-Learning

Initially, in being involved with the process of m-Learning, one should admit the basic elements of such a process, such as the teacher, the learner, the surrounding environment as well as the contents and, finally, assessment. Therefore, to simplify the matter, the following Fig. 1 shows such an easy involvement.

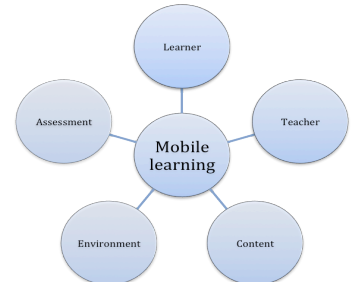

Adapted from: Ozdamli and Cavus bas [8]

Figure 1. The basic elements of the m-Learning process

Mobile technologies have already demonstrated in many countries that one can get benefits from using such devices in the education process [8]. m-Learning development offers endless valuable opportunities to both students and learners, considering the emergence of the highest technology of both networks and mobile devices [2]. Hence, we can assert that mobile technologies, especially after many continued years of evolving and development, have become so mature, simply to support the process of learning. The involved advanced devices of mobiles, for example, PDAs, iPod, wireless network and mobile phones techniques, might assist in making the m-Learning process more feasible, especially in teaching and learning [8]. Moreover, there is a new attitude among the educational experts and institutions for adopting m-Learning, simply as such instructional strategies. Hence, one can forecast such alternative modes related to m-Learning concerning the involved learners' education, whereas such techniques allow more flexibility of the process of learning [1].

The matter became easy to employ such technique successfully, whereas the usage of m-Learning could verify the possibility of catching up with such ever changing world and moreover, to also boost collaboration and interaction among teachers and students [8].

Figuring out the starting ages of students for employing mobile techniques could indicate that new generations could already be adapted to this technology, simply to use mobile techniques in supporting the learning in classrooms. Such new generations are living in an environment that is full of mobile technologies [2]. Therefore, when one could use the mobile techniques regularly, the highest quality would be expected, simply in order to be boosted up later. m-Learning, consequently, would be our learning future that logically could expand the process of E-learning and, additionally, could have such potential to further expansion that, in turn, can make the process of learning available for all [3].

On the other hand, while developing the process of m-Learning, one should consider the new mode of learning with care, particularly when implementing it as a new educational option for learners. The matter requires balance, especially with the needs of students and, as well with such rapid technical development [4].

## B. Advantages of m-Learning

Indeed, m-Learning involves many advantages, such as making the learning process easy at any time and at any place. This can save times and efforts of teachers, whereas the process of education can be involved with fun. In order for the m-Learning be successful and to obtain the highest benefits and advantages from it, one should consider the correspondent characteristics of such a valuable process [8].

Fig. 2 shows the characteristics of m-Learning;

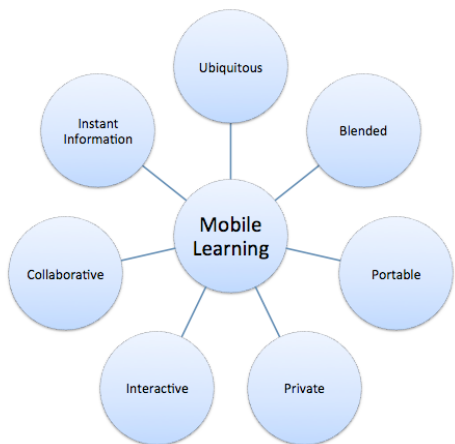

Adapted from: Ozdamli and Cavus [8]

Figure 2. The characteristics of m-Learning

Ozdamli and Cavus [8] specified many characteristics of m-Learning, such as (i) ubiquitous/ spontaneous, (ii) the tools of the mobiles' portable size, (iii) private, (iv) blended, (v) collaborative and interactive, and (vi) instant information.

## C. Disadvantages of m-Learning

Besides several advantages of m-Learning, one should consider the involved disadvantages of using m-Learning. Indeed, there are the set up costs for the involved equipment acquisition and, additionally, such training costs in the educational organization [11]. Moreover, the facts of copyright and security issues have recently become the main concern of the involved institutions; this may, indeed, expose such valuable data to unauthorized individuals, if the mobile technologies used appear insecure or even unsafe [12].

However, the matter still requires admitting that m-Learning is a creative and valuable means by which one can enhance the experience of learning, either at the workplace or in education [1] [12].

## D. Challenges in m-Learning

Vavoula and Sharples [13] specified six challenges in evaluating m-Learning: analyzing and capturing learning in or across context, measuring the process and outcomes of m-Learning, respecting the privacy of the learner/ participant, assessing the utility or even usability of mobile devices, regarding the wider context of an organization, the socio-culture of learning, and additionally, informality evaluation.

However, the authors admitted that those six challenges result from the complicated nature of learning based on mobile devices, where the focus was social, instead of being technical [13].

## E. Difference between e-Learning and m-Learning

Initially, it is easy to notice that the existing difference between m-Learning and E-Learning can be found in the differing capabilities of the web browsers involved in each environment. However, it may be tempting to see E-Learning as an alternative to campus or classroom learning; therefore, m-Learning can take such learning away, especially from certain fixed point. Whereas E-Learning is considered an alternative to learning in classrooms, m-Learning can be a complementary activity, especially to both traditional learning and E-Learning. Most importantly, m-Learning considers the fact that the user involved will interact with many educational resources, while the students are apart from the normal place of education, the computer or even the classroom [7] [14] [15].

E-Learning has, indeed, been around for many more years than m-Learning, particularly with many types of the recent employment of m-Learning. Critically, m-Learning has been winning great approval, especially among young students who have grown up employing their portable videogames as well as their highest wireless technological devices. Corresponding to such meaning, m-Learning can appeal, not only to the persons who actually require portable, learning, but also to others who would like to use their mobile devices in either learning or playing games [7] [15].

## F. Review of e-Learning maturity models

In spite of the adoption of E-Learning, which was demonstrated rapidly among most of those education institutions, one should ensure that there is an effectiveness and efficiency in the employment of such techniques for the organizational resources that may appear to be hard tasks. In literature there are many resources that handled such techniques with care [2] [4]. Corresponding to this point, one can figure out that many challenges appear simply because the many users involved cannot employ the convenient software and understanding the usage of such techniques [16].

What is actually required is to find out such a process model by which the users will be able to get enough flexibility to guide the enhancement of the process of m-Learning. Along with the development of such a technique, the capability maturity models (CMM) appear simply to solve such challenges. CMM is a 5-level model that helps to judge the maturity of the software used in the institutions or organizations and, moreover, to identify the critical steps and other validated practices needed for the effectiveness or capability of the process.

Five levels of such model can be figured out as the following [10]:

*1) Initial:* in this case the developed process can be characterized as ad-hoc, whereas little processes can be defined and the resulting success may focus basically on the persons' heroics and efforts.

*2) Repeatable:* the process of the main project management is to be established to a certain track schedule, cost, as well as functionality. The process discipline needed can be seen in its place, in order to repeat the earlier success on such projects, especially with same application.

*3) Defined:* the development and management activities can be standardized, documented, and integrated into a set of friendly standard processes for the institute or organization involved.

*4) Managed:* the detailed process measures as well as the quality of products can be collected to make the process easy and the product involved can be controlled and understood.

*5) Optimizing:* the regular and continued process enhancement can be facilitated, particularly by feedback from the process involved, as well as from the piloting technologies and other innovative ideas.

Critically, CMM's originality was designed in order to offer these various benefits, such as offering certain road maps for enhancing the software development process of the institute or organization [10].

## V. RELATED WORK

Indeed, one can figure out that the basic theme of CMM, particularly for the process of education, cannot be considered a new concept as mentioned by [17].

Checking E-Learning Maturity Model (EMM), one can refer to the argument of Marshall & Mitchell [18][19][20] that focused on the enhancement of such a process of the involved software and, additionally, the ability of the model dEtermination SPICE ISO/IEC [16]. However, one can still notice that SPICE is considered a reply of ISO, especially on the CMM of SEI', whereas it could employ mainly those 5 levels of the maturity, added to extra standard zero; indeed, for such process that could not be performed, or even might be performed on an incomplete basis.

Truly, the best practices and benchmarks for the process of E-Learning were examined and employed to compile a certain set of the targeted practices that might be suitable in 5 areas. The levels of maturity, unlike CMM, could be employed mainly in evaluating each of the practices, not only for the entire organization. However, one may still notice that the given results can be a framework, obviously for evaluating the maturity of E-Learning that could be applied to the systems of E-Learning of many universities in New Zealand [18][19].

It appears that the basic objectives of EMM regarding the educational process may appear like those of CMM, whereas its revealed domain appears different. This is not convenient to our aim. There are certain practices that could be applied, such as Neuhauser [21] offered OCDMM (Online Course Design Maturity model that as well may be considered as E-Learning's maturity model, focusing on CMM. Such a model can describe various states of E-learning technologies' adoption. However, mainly the difference between those levels can be figured out in the degree to which the technology of E-Learning can be employed successfully.

According to Kajko-Mattsson [22], there is a proposal of $CM^3$, (Corrective Maintenance Maturity Model), especially for maintainers training and education. Such s model is, indeed, focusing on two main educational processes, especially from the industry involved and various process models generically. Among them, one can figure out CMM. Moreover, it could define 3 main levels of the corresponded maturity: primary, defined, and optimal. Indeed, it is easy to admit that such a maturity model aims toward continued education, especially in the correspondence environmental industry that is radically different from the educational organization one. Unlike our specified domain, one can notice that education, indeed, is not as such the basic focus of $CM^3$'s context. Consequently, despite the fact that both maturity models are corresponding to education, one still admits that they are different in full, and we cannot compare them [17].

However, the CMM model has been successfully modified and adapted to assess maturity in varied software application domains like the usability of open source software [23] and software product line engineering maturity [24]. This proves that with appropriate modifications, the CMM model can be adapted to assess the maturity of m-Learning within educational institutions as well, which is the primary goal of this paper.

To sum up, one can figure out that CMM, therefore, has not at all been applied critically on the domain of teaching, generally, except for the above-mentioned work [17]. As clarified later, one can recognize that the other educational maturity models appear to either rely on developing professionalism in the context of industry or on the process of E-Learning, considered a special domain. Still one can urge that neither of such approaches could be applicable on the m-Learning generally, yet may later inspire the best analogous practices. Indeed, from conducting such analysis one can see that m-Learning modules, coupled with certain structured assessments, have their potential simply in order to improve the educational experiences.

## VI. PROPOSED M-LEARNING MATURITY MODEL

From such critical point, the main target therefore is to figure out a proposed model for the m-Learning maturity model. The authors have published a research paper [25] as an initial maturity model for m-Learning. In this paper a new level were add to there are six basic stages in the proposed model as follows:

*Level 0:*

Based on the SPICE ISO/IEC [16] level 0 in this case it is recognized as a limited mobile presence. In this stage, universities and the institutions do not consider mobile devices to be important in the provision of their services and products. As a result, there are no systems or procedures to encourage the students to utilize the mobile devices; therefore, they do not invest in developing mobile learning systems.

*Level 1:*

This level is known as the preliminary stage. It is a reactive and experimental stage that, once initiated by the learning institution, recognizes the need to improve the provision of information to students through mobile devices. The institutions react to the external pressures and needs for flexibility and convenience by students.

In this level, the institution has the pilot program for implementation but there is a lack of a vision to guide the implementation. The institutions develop measures to facilitate the implementation of the prototypes. This is done experimentally but is hampered for a number of reasons. For instance, the mobile device coverage might be limited or students might not understand the value of the mobile learning environment.

Another limitation in the implementation of the prototype might be the fact that the learning institution might not have the ability to facilitate effective implementation. In the preliminary stage, most of the universities and institutions do not have clear mobile learning policies and defined objectives to guide mobile learning; these factors might limit the implementations and the usage of the systems.

*Level 2:*

The establishment stage, it is based on the recognition of the opportunity provided by mobile devices in the education system. This resulted in the investment of m-Learning technologies to realize the opportunities provided. In this stage, learning institutions formulate clear objectives to guide m-Learning implementation. However, those Institutions do not have m-Learning mechanisms to evaluate their systems. This brings the need for improvements on the existing and implemented pilot prototypes. Programmers develop systems to facilitate the use of mobile learning in education such as the Android App Education and iOS App Education[3].

*Level 3:*

In the defined stage, the model of mobile learning environment has been developed to measure the quality of mobile learning systems. The focus on learning mobile systems by institutions features to offer the most mobile platforms. In the defined stage, the mobile device is considered as a critical tool in the interaction between students or among students, instructors and administrative staff. Learning institutions must link their mobile learning strategies with core and technical visions, and they must invest heavily in this type of systems to achieve success. This can be done by establishing clear guidelines to reach the desired level of success.

*Level 4:*

The structured stage; in this level, m-Learning is characterized by optimization and innovation. The optimization results in a rich, dynamic, and flawless experience for students and tutors in the use of the system. To solidify their systems, institutions borrow and integrate the best practices from other institutions. Universities develop and measure to ensure a real time student engagement and context awareness. Also, they develop systems to be used in different mobile devices such as tablets and mobile phones. The use of mobile device applications allows students to provide feedback, give comments, and share information. As a result, institutions refine and improve procedures and policies to control any changes experienced in mobile changes.

*Level 5:*

Finally, the continuous improvement stage; in this stage mobile offering has already been accepted as the best approach to provide knowledge and exchange of information between students and instructors. In this stage, institutions are constantly evaluating themselves to ensure continuous improvement and optimization. This helps identify any changes that occur that might limit or change the manner in which mobile learning is used.

## VII. CONCLUSION

Getting involved with the highest technical tools may be confusing, especially when they are used in the education process. Checking the differences between m-Learning and E-Learning, one can understand that the first is most recently developed, especially after the high revolution in creating mobile devices, which can be used widely by either students or others. In turn, m-Learning can make the connection easier, since it does not need to be limited into certain place, such as computers.

This research contributes toward an initial maturity evaluation of mobile learning. However, the maturity model illustrated helps understand the realization of an m-Learning model. It provides an assessment of the provision of education through mobile learning and the capabilities of the method. It also provides a literature review on m-Learning, helping show its effectiveness and challenges.

The value gained from such a paper is demonstrating the possibility of adapting CMM in order to present such a valuable road map especially to support institutional and individual efforts that might enhance the process of organization that could be collaborated with m-Learning. If we encounter such questions, whereas it is a valid process, we can simply answer that while it is obvious that such models appear incomplete, they can, however, present many benefits.

In short, one can add that such mentioned frameworks could not provide a complete of such list of the necessary key processes, particularly for enhancing the process of m-Learning, just such indicative set of those possible results, in order to define each of those levels that may be possible in the maturity model. This is not an exhaustive study, but it still contributes many possible results. Clearly, the following step in developing such a model could be figured out in identifying the key activities that have led to the improvements in earning m-Learning.

We hope in the future to conduct a comprehensive study that will involve empirically testing the maturity model illustrated and help to develop a statistical model on the same.

ACKNOWLEDGMENT

The first author would like to thank the Ministry of Higher Education (MOHE) in Saudi Arabia for his scholarship.